\documentstyle[twoside,12pt,epsf,amssymb]{article}
\begin{document}
\title{\bf Construction of 
particular solutions to 
nonlinear equations of Mathematical Physics by using matrix
algebraic equation}

\author{ Alexandre I.
Zenchuk
\\Center of Nonlinear Studies of\\ L.D.Landau Institute 
for Theoretical Physics  \\
(International Institute of Nonlinear Science)\\
Kosygina 2, Moscow, Russia 117334\\
E-mail: zenchuk@itp.ac.ru
\\}
\maketitle

\begin{abstract}
The paper  develops the method for construction of the families of
particular solutions to the nonlinear Partial Differential
Equations (PDE) without relation to the complete integrability. Method
is based on the specific link between algebraic matrix equations
and PDE.
Example of (2+2)-dimensional generalization of Burgers equation
is given.
\end{abstract}

\noindent
PACS numbers:  05.45.-a, 05.45.Yv

\noindent
Keywords: nonlinear Partial Differeential Equations, kink,
integrability, Burgers equation

\section{Introduction}
Analysis  of nonlinear Partial Differential Equations (PDE) is
severe problem in mathematical physics.  Many different methods have
been developed for analytical investigation of nonlinear PDE
during last decades:
Inverse
Scattering Problem \cite{GGKM,ZSh1,ZSh2,ZM1,M,ZMNP,AKNS},
Sato theory \cite{OSTT,PM,P,Z4}, Hirota bilinear method \cite{H1,H2,HS,HS2}, 
Penlev\'e method \cite{WTC,W,EG}, 
$\bar{\partial}$-problem \cite{ZM,BM,K}, with some generalizations
\cite{DMZ,Z5,Z6,KZ,Z7},
nonlinear $\bar{\partial}$-problem \cite{KMR,BKM,KMM}, recent modification 
of the dressing method 
based on the algebraic matrix equation \cite{Z1,Z2,Z3}. 
A wide class of PDE  (so-called completely
integrable systems) has been studied better then others. 
Nevertheless, there are many methods which work in nonintegrable
case as well:  \cite{H1,H2,HS,HS2,WTC,W,EG,Z1,Z2,Z3}.  

We represent the method for construction of families of
particular solutions to 
multidimensional nonlinear PDE
 without relation to the complete integrability. This method is
 based on general properties of linear algebraic matrix
 equations. Essentially we 
  develop some ideas represented in the ref.\cite{OSTT}  and
recently in the ref. \cite{Z4}.  

General algorithm is discussed in the Sec. 2. Sec. 3 deals with
example of (2+2)-dimensional  system of PDE together with family of particular
solutions. By using an appropriate reduction, this system 
leads to Burgers equation.
This example is simple demonstration of the method.
Conclusions are given in the Sec.4.

\section{General results}

The algorithm represented in this section is based on the
fundamental properties of linear matrix algebraic equation, which
is written in the following form:
\begin{eqnarray}\label{MATR}
\Psi U = \Phi,
\end{eqnarray}
where $\Psi=\{\psi_{ij}\}$ is $N\times N$  matrix
$U=[u_1,\dots,u_N]^T$, $\Phi=[\phi_1,\dots,\phi_N]^T$.
Let us recall these properties.

\begin{enumerate}
\item
 If $\Psi $ is nondegenerate matrix, i.e.
\begin{eqnarray}\label{DET}
\det \Psi\neq 0,
\end{eqnarray}
then equation (\ref{MATR}) has unique solution which can be written in the
following form:
\begin{eqnarray}\label{SOL}
U=\Psi^{-1} \Phi.
\end{eqnarray}
Only nondegenerate matrices $\Psi$ will be considered hereafter.
\item
({\it consequence of the previous property}) 
If $\Phi=0$ and condition (\ref{DET}) is held, then the 
equation (\ref{MATR}) has only the trivial solution
\begin{eqnarray}\label{SOLZ}
U\equiv 0.
\end{eqnarray}
\item
 ({\it superposition principle}) 
Consider the set of $K$ matrix equations with the same matrix $\Psi$:
\begin{eqnarray}\label{SUPER}
\Psi U_i = \Phi_i,\;\;i=1,\dots,K.
\end{eqnarray}
Then for any set of scalars $b_k$ ($k=1,\dots,K$),
 function $\tilde U=\sum_{k=1}^K b_k U_k$
is solution of the following matrix equation
\begin{eqnarray}\label{MATRSUPER}
\Psi \tilde U = \sum_{k=1}^K b_k\Phi_k.
\end{eqnarray}
\item
 ({\it consequence of properties 2 and 3})
If columns $\Phi_i$ are linearly dependent,
i.e there are scalars $a_k$, $k=1,\dots,K$, such that
\begin{eqnarray}\label{FUND1}
\sum_{k=1}^K a_k \Phi_k =0,\;\; 
\end{eqnarray} 
then 
\begin{eqnarray}\label{FUND2}
\sum_{i=1}^K a_i U_i  =0.
\end{eqnarray}
\end{enumerate}
Note that  analogous properties of linear integral equation have been
used in the 
classical dressing method based on the $\bar{\partial}$-problem
\cite{ZM,BM}.

To relate the matrix equation (\ref{MATR}) with the system of nonlinear
PDE, one needs to introduce the set of additional variables 
${\bf x}=(x_1,\dots,x_M)$ in matrices $\Psi$, $\Phi$ and $U$, 
which are independent variables of
succeeding system of nonlinear PDE. $M$ is dimension of ${\bf
x}$-space. For this purpose we use the following  system of
linear differential 
equations
\begin{eqnarray}\label{VAR}
M_{k}\Phi + N_k\Psi =\Psi A_k,\;\;k=1,\dots,P,
\end{eqnarray}
where $M_k$ and $N_k$ are linear differential operators with 
matrix coefficients;
 $A_k$ are $N \times N$ matrices, which are constant for the sake of
 simplicity. The number of these equations, $P$, depends on
 situation.
 Equations (\ref{VAR})
introduce variables ${\bf x}$ into the matrices $\Psi$ and $\Phi$.
Due to the eq.(\ref{SOL}),  elements
$u_i$ of the column $U$ are functions of variables ${\bf x}$ as
well.

Operators $M_k$ and $N_k$ may not be arbitrary. They
have to satisfy two conditions:
\begin{enumerate}
\item
 Overdetermined system (\ref{VAR}) is compatible.
\item
Operators $M_k$ and $N_k$ have to   provide   
existence of  differential  operator ${\cal{M}}$ 
with {\it non-constant} matrix coefficients such that
\begin{eqnarray}
{\cal{M}} (\Phi - \Psi W)&=& \sum_{k=1}^P (M_k \Phi + N_k \Psi)\tilde U_k
+ \Psi \tilde V = \Psi B(U) \equiv 0,\;\;\\\nonumber
B(U)&=&\sum_{k=1}^P A_k \tilde U_k + \tilde
V ,
\end{eqnarray}
where $\tilde U_k $ and $\tilde V$  are 
some matrix functions of elements $u_i$ and their
derivatives (see eqs.(\ref{phi0}-\ref{phiL}) as an example).
\end{enumerate}
Since ${\mbox{det}}\Psi\neq 0$, the last equation means: 
\begin{eqnarray}
B(U)=0,
\end{eqnarray}
which represents the system of nonlinear PDE on functions
$u_i$. The structure of this system is defined by the equations
(\ref{VAR}).

One can see that the matrix equation (\ref{MATR})
is generalization of the linear systems used in the
refs.\cite{OSTT,Z4}.
In fact, let us introduce notations 
\begin{eqnarray}
\phi_i \equiv \phi^{(i)},\;\;\psi_{ij}\equiv \psi_j^{(i)},
\end{eqnarray}
and rewrite the matrix equation (\ref{MATR}) in the form of system of $N$
equations:
\begin{eqnarray}\label{ALGSYST}
\phi^{(i)} = \sum_{j=1}^N u_j \psi_j^{(i)},\;\;i=1,\dots,N.
\end{eqnarray}
Two evident reductions are possible (these reductions will not be
used in the succeeding sections).

1.
If 
\begin{eqnarray}
\psi_j^{(i)} = \partial_{x_1}^j \psi^{(i)},\;\; \phi^{(i)}
=\partial_{x_1}^{(N+1)} \psi^{(i)},\;\; i,j=1,\dots,N,
\end{eqnarray}
then system (\ref{ALGSYST}) is system of $N$ ordinary differential equations on
the function $\psi^{(i)}$ which is basic for the algorithm
represented in the
ref.\cite{OSTT}.

2. 
Introduce multi-index 
$\beta =(\beta_1, \dots, \beta_M)$, $||\beta||=\sum_{k=1}^M
\beta_k$ instead of index $j$ ($M$ is dimension of ${\bf x}$-space),  
and use notations
\begin{eqnarray}
\psi^{(i)}_\beta &=& \prod_{k=1}^M
\partial_{x_k}^{\beta_k}\psi^{(i)} \equiv \partial^\beta
\psi^{(i)},\;\;||\beta||=1,\dots,N,\;\;\\
 \phi^{(i)}&=& 
\partial^\alpha \psi^{(i)},\;\;||\alpha||=N+1.
\end{eqnarray}
Then the system (\ref{ALGSYST}) can be written in the form
\begin{eqnarray}\label{ALPHA}
\partial^\alpha \psi^{(i)} = \sum_{||\beta||=0}^N 
u_\beta \partial^\beta \psi^{(i)},\;\; i=1,\dots,Q,
\end{eqnarray}
where $Q$ is the number of terms in the right hand side of the
eq.(\ref{ALPHA}), $\alpha$ takes any value such that $||\alpha||=N+1$.
Eq.(\ref{ALPHA}) is basic in the ref.\cite{Z4}.

Hereafter,  we will use eq. (\ref{ALGSYST}) 
instead of matrix equation (\ref{MATR}).
Superscript $i$ runs the values from 1 to $N$, unless otherwise
specified.

\section{Example: (2+2)-dimensional generalization of Burgers
equation}

In this section we consider (2+2)-dimensional equations with
following notations for independent variables:
\begin{eqnarray}\label{var}
x\equiv x_1,\;\;y\equiv x_2, \;\;t\equiv x_3,\;\;\tau\equiv x_4.
\end{eqnarray}
For convenience of construction, we separate the first term of
the sum in the eq. (\ref{ALGSYST}):
\begin{eqnarray}\label{ALGSYSTN}
\phi^{(i)}=u_1 \psi_{1}^{(i)} +\sum_{k=2}^N u_k \psi_k^{(i)}.
\end{eqnarray}
Introduce two differential operators $L_1$ and $L_2$:
\begin{eqnarray}\label{LL}
L_1&=&l_1 +\delta_{11} {u_1}_x + \delta_{12} {u_1}_y,\;\;
l_1=\partial_t + \partial_x^2 + \gamma_1 \partial_{x}\partial_y,
\\\nonumber
L_2&=&l_2 +\delta_{21} {u_1}_x + \delta_{22} {u_1}_y,\;\;
l_2=\partial_\tau + \partial_y^2 + \gamma_2 \partial_{x}\partial_y.
\end{eqnarray}
Let us apply operator $L_1$ to the eq.(\ref{ALGSYSTN}) and 
investigate the result
\begin{eqnarray}\label{APPLL1}
-l_1\phi^{(i)} -\delta_{11} {u_1}_x \phi^{(i)} - \delta_{12} {u_1}_y \phi^{(i)} +
(L_1 u_1) \psi_1^{(i)} + u_1 (l_1 \psi_1^{(i)}) + &&\\\nonumber
 {u_1}_x (2 {\psi_1^{(i)}}_x +
\gamma_1 {\psi_1^{(i)}}_y)+ 
\gamma_1 {u_1}_y  {\psi_1^{(i)}}_x +&&\\\nonumber
\sum_{k=1}^N \left(
(L_1 u_k) \psi_k^{(i)} + u_k (l_1 \psi_k^{(i)}) +  {u_k}_x (2 {\psi_k^{(i)}}_x +
\gamma_1 {\psi_k^{(i)}}_y)+ \gamma_1 {u_k}_y  {\psi_k^{(i)}}_x 
\right)&=&0
\end{eqnarray}
Our purpose is to represent the equation (\ref{APPLL1}) in the form of linear
homogeneous equation
\begin{eqnarray}\label{F1k}
\sum_{j=1}^N F_{1k} \psi_k^{(i)} =0,
\end{eqnarray}
where $F_{1k}$ are some expressions of $u_k$ and their derivatives.
Owing to the property 4 (eqs.(\ref{FUND1}) and (\ref{FUND2})), 
this would mean that
$F_{1k}\equiv 0$, $k=1,\dots,N$.
The simplest way to result in the eq. (\ref{F1k}) is introduction
of the  system of differential equations (\ref{VAR}) on the functions 
$\psi_j^{(i)}$ and $\phi^{(i)}$, which looks as follows:
\begin{eqnarray}\label{phi0}
\phi^{(i)}-\frac{1}{\delta_{11}} (2 {\psi_1^{(i)}}_x +
\gamma_1 {\psi_1^{(i)}}_y) &=& \sum_{k=2}^N \tilde \alpha_{1k}
\psi_k^{(i)},\;\;\\\label{phi1}
\phi^{(i)}-\frac{\gamma_1}{\delta_{12}} 
{\psi_1^{(i)}}_x &=& \sum_{k=2}^N  \alpha_{1k} \psi_k^{(i)},
\end{eqnarray}
\begin{eqnarray}
\label{psi2}
{\psi_2^{(i)}}_x &=& \beta_1 \psi_1^{(i)} + \sum_{k=2}^N \alpha_{3 k}
\psi_k^{(i)},\;\;
{\psi_2^{(i)}}_y \;=\; \beta_2 \psi_1^{(i)} + \sum_{k=2}^N \alpha_{4 k} \psi_k^{(i)},\\
\label{psij}
{\psi_j^{(i)}}_x &=& \sum_{k=2}^N \alpha_{(2j-1) k} \psi_k^{(i)},\;\;
{\psi_j^{(i)}}_y \;= \;\sum_{k=2}^N \alpha_{(2j) k} \psi_k^{(i)},\;\;j>2,\\
\label{psiL}
l_1{\psi_j^{(i)}} &=& \sum_{k=2}^N \beta_{1jk} \psi_k^{(i)},\;\;
l_2{\psi_j^{(i)}}\;=\;\sum_{k=2}^N \beta_{2jk} \psi_k^{(i)},\;\;j\ge 1,\\
\label{phiL}
l_1{\phi^{(i)}} &=& \sum_{k=2}^N \beta_{10k} \psi_k^{(i)},\;\;
l_2{\phi^{(i)}} \;=\; \sum_{k=2}^N \beta_{20k} \psi_k^{(i)}.
\end{eqnarray}
Here parameters 
$\tilde \alpha_{ij},\alpha_{ij}, \beta_i, \beta_{ijk}$ 
have to provide compatibility of the overdetermined system
(\ref{phi0})-(\ref{phiL}). 
Equation (\ref{phi0}) can be simplified if one eliminates
$\phi^{(i)}$  by using
the eq.(\ref{phi1}):
\begin{eqnarray}\label{psi1}
{\psi_1^{(i)}}_y-G {\psi_1^{(i)}}_x =  \sum_{k=2}^N \alpha_{2 k}
\psi_k^{(i)},\;\;
G=  \frac{\delta_{11}}{\delta_{12}}-\frac{2}{\gamma_1} ,\;\;
\alpha_{2k}= \frac{\delta_{11}}{\gamma_1}
(\alpha_{1k}-\tilde\alpha_{1k}).
\end{eqnarray}
The system (\ref{phi1})-(\ref{psi1}) will be used below.

Now the equation (\ref{APPLL1}) takes the form (\ref{F1k}) with
\begin{eqnarray}\label{eq1}
F_{11}={u_1}_t + {u_1}_{xx} + \gamma_1 {u_1}_{xy} + 
\delta_{11} u_1 {u_1}_x+\delta_{12} {u_1} {u_1}_y +&& \\\nonumber
(2 \beta_1+ \beta_2 \gamma_1) {u_2}_x +
 \beta_1 \gamma_{1} {u_2}_y&=&0.
 \end{eqnarray}
Expressions for $F_{1k}$ with $k>1$  will not be used.

 Analogously,
 apply operator $L_2$ to the eq.(\ref{ALGSYSTN}):
 \begin{eqnarray}
-l_2\phi^{(i)} -\delta_{21} {u_1}_x \phi^{(i)} - \delta_{22} {u_1}_y \phi^{(i)} +
(L_2 u_1) \psi_1^{(i)} + u_1 (l_2 \psi_1^{(i)}) + && \\\nonumber 
{u_1}_y (2 {\psi_1^{(i)}}_y +
\gamma_2 {\psi_1^{(i)}}_x)+
\gamma_2 {u_1}_x  {\psi_1^{(i)}}_y +&& \\\nonumber
\sum_{k=1}^N \left(
(L_2 u_k) \psi_k^{(i)} + u_k (l_2 \psi_k^{(i)}) +  {u_k}_y (2 {\psi_k^{(i)}}_y +
\gamma_2 {\psi_k^{(i)}}_x)+ \gamma_2 {u_k}_x  {\psi_k^{(i)}}_y 
\right)&=&0
\end{eqnarray}
 Due to the relations (\ref{phi1})-(\ref{psi1}) one gets analogous expression
 \begin{eqnarray}\label{F2k}
\sum_{j=1}^N F_{2k} \psi_k^{(i)} =0,
\end{eqnarray}
 if only
 \begin{eqnarray}
 \delta_{21}=\frac{\gamma_2}{\gamma_1^2}( \delta_{11}
 \gamma_1-2\delta_{12}),\;\;
 \delta_{22}=\frac{1}{\gamma_1^2}( \delta_{12}
 \gamma_1\gamma_2+2\delta_{11}\gamma_1 - 4 \delta_{12}).
 \end{eqnarray}
 Coefficients $F_{2k}$ are expressed in terms of the functions
 $u_k$ and their derivatives with
 \begin{eqnarray}\label{eq2}
F_{21}={u_1}_\tau+{u_1}_{yy}+ \gamma_2 {u_1}_{xy} +
\frac{\gamma_2}{\gamma_1^2}(\delta_{11}\gamma_1-2
 \delta_{12})u_1 {u_1}_x+&&\\\nonumber
\frac{1}{\gamma_1^2} (\delta_{12} \gamma_1\gamma_2 + 2
\delta_{11}\gamma_1 -4 \delta_{12}) u_1 {u_1}_y+
\beta_2 \gamma_{2} {u_2}_x+
(2\beta_2+\beta_1\gamma_2 ) {u_2}_y&=&0,
\end{eqnarray}
Eqs.(\ref{eq1}) and (\ref{eq2}) form  the system of two nonlinear PDE on
the functions $u_1$ and $u_2$.

One shell discuss the structure of the system
(\ref{phi1}-\ref{psi1}).
Operators $L_1$ and $L_2$, given by the formulas  (\ref{LL}),
determine the left hand sides of the equations in this system. 
The structure of the system of nonlinear PDE is defined 
by the operators $L_1$ and $L_2$ and by the right hand sides of the
equations in the system (\ref{phi1}-\ref{psi1}). Only terms
proportional to $\psi_1^{(i)}$ effect on the structure
of the
nonlinear system of PDE.  At least one of the equations (\ref{phi1}-\ref{psi1})
must have this term (see eqs.(\ref{psi2})). Otherwise, only
simplest solutions to the nonlinear system of PDE are available.
Other terms enrich the family of available
particular solutions.

Function $\phi^{(i)}$ may not be linear combination of functions
$\psi_j^{(i)}$ (see eqs. (\ref{phi0},\ref{phi1})).  Otherwise all
coefficients $u_j$ in eq.(\ref{ALGSYSTN}) are zero.
Equations (\ref{psiL}) and (\ref{phiL})  introduce dependence on 
variables $t$ and $\tau$. Right hand sides of these equations are
zero  in the
examples below for the sake of simplicity.

\subsection{Construction of  particular solutions}

This section is devoted to construction of some families of
particular solutions to the system (\ref{eq1},\ref{eq2}).
We will use simplified form of the system (\ref{phi1}-\ref{psi1}):
\begin{eqnarray}\label{phi1n}
\phi^{(i)}&=&\frac{\gamma_1}{\delta_{12}}
{\psi_1^{(i)}}_x +\alpha_{12} \psi_2^{(i)},\\\label{psi1n}
{\psi_1^{(i)}}_y&=&G{\psi_1^{(i)}}_x ,\;\;
G=  \frac{\delta_{11}}{\delta_{12}}-\frac{2}{\gamma_1} ,
\\\label{psi2n}
{\psi_2^{(i)}}_x&=&\beta_1\psi_1^{(i)}+\alpha_{32} \psi_2^{(i)},\;\;
{\psi_2^{(i)}}_y\;=\;\beta_2\psi_1^{(i)}+\alpha_{42} \psi_2^{(i)},\\\label{psijn}
{\psi_j^{(i)}}_x &=& \sum_{k=2}^N \alpha_{(2j-1) k} \psi_k^{(i)},\;\;
{\psi_j^{(i)}}_y \;=\; \sum_{k=2}^N \alpha_{(2j) k}
\psi_k^{(i)},\;\;j>2,
\\\label{psiphiLn}
l_k\psi_j^{(i)}&=&0,\;\;l_k\phi^{(i)}\;=\;0,\;\;k=1,2,\;\;j=1,\dots,N
\end{eqnarray}
Below are two examples with $N=2$ and $N=4$. Example with $N=3$
does not have a principal difference with the case  $N=2$.
Solutions from both 
families  considered below are
parameterized
by the 
arbitrary functions of single variable.

\subsubsection{One-dimensional kink}

Let $N=2$, superscript $i$ takes the values 1 and 2. 
The system (\ref{ALGSYSTN}) has the following form:
\begin{eqnarray}\label{algsyst1}
\phi^{(i)}&=& u_1 \psi_1^{(i)} + u_2 \psi_2^{(i)},
\end{eqnarray}
with
\begin{eqnarray}
\label{sol1}
u_1&=&\frac{\triangle_1}{\triangle},\;\;u_2\;=\;\frac{\triangle_2}{\triangle},
\end{eqnarray}
\begin{eqnarray}
\label{det1}
\triangle=\left|\begin{array}{cc}
\psi_1^{(1)} & \psi_2^{(1)}\cr
\psi_1^{(2)} & \psi_2^{(2)}\end{array}
\right|,\;\;
\triangle_1=\left|\begin{array}{cc}
\phi^{(1)} & \psi_2^{(1)}\cr
\phi^{(2)} & \psi_2^{(2)}\end{array}
\right|,\;\;
\triangle_2=\left|\begin{array}{cc}
\psi_1^{(1)} & \phi^{(1)}\cr
\psi_1^{(2)} & \phi^{(2)}\end{array}
\right|,\;\;
\end{eqnarray}
We need only  equations (\ref{phi1n})-(\ref{psi2n}) and
(\ref{psiphiLn}) (with $j=1,2$) to find functions
$\psi_j^{(i)}$ ($j=1,2$) and $\phi^{(i)}$.
They admit the following solutions 
\begin{eqnarray}\label{solpsi1}
\psi_1^{(i)}&=&\int c_1^{(i)}(k) e^{k x+ q y + \omega t +\nu \tau} d k,
\\\label{solpsi2}
\psi_2^{(i)}&=&\int c_2^{(i)}(k) e^{k x+ q y + \omega t +\nu \tau} d k,
\\\label{solphi}
\phi^{(i)}&=&\int c^{(i)}(k) e^{k x+ q y + \omega t +\nu \tau} d k,\\
\label{ci}
c^{(i)}(k)&=&\frac{c_2^{(i)}(k)}{\beta_1\delta_{12}} (
\gamma_1 k^2-
\alpha_{32} \gamma_1 k +\alpha_{12} \beta_1 \delta_{12} ),\\\label{c1i}
c_1^{(i)}(k)&=&\frac{c_2^{(i)}(k)}{\beta_1}(k-\alpha_{32}),\\
\label{q}
q&=&\frac{\beta_2}{\beta_1} k,\;\;\\\label{omega}
\omega&=&-k^2 - k q\gamma_1=-\frac{k^2}{\beta_1}(\beta_1+\beta_2 \gamma_1),
\\ \label{nu}
\nu&=&-q^2 - k q\gamma_2=-\frac{k^2
\beta_2}{\beta_1^2}(\beta_2+\beta_1 \gamma_2),\\
\label{param1}
\delta_{11}&=&\frac{\delta_{12}}{\beta_1\gamma_1} (\beta_2 \gamma_1
+ 2 \beta_1),\;\;\alpha_{42}\;=\;\frac{\beta_2}{\beta_1} \alpha_{32}
\end{eqnarray}
One can see from the eqs.(\ref{solpsi1}-\ref{c1i}) that functions
$\psi^{(i)}_2$ are arbitrary functions
of single variable (say, variable $x$). 
So that appropriate solutions (\ref{sol1}) depend on two arbitrary functions
of single variable.

As a simple example, let us take the following expressions for
the functions $\psi_j^{(i)}$ and $\phi^{(i)}$:
\begin{eqnarray}
\psi_1^{(1)}&=&\frac{k_1-\alpha_{32}}{\beta_1}s_1 E_1,\\
\psi_1^{(2)}&=&\frac{k_2-\alpha_{32}}{\beta_1}s_2 E_2+
\frac{k_3-\alpha_{32}}{\beta_1}s_3 E_3,\\
\psi_2^{(1)}&=&s_1 E_1,\\
\psi_2^{(2)}&=&s_2 E_2+s_3 E_3,\\
\phi^{(1)}&=&\frac{s_1E_1}{\beta_1\delta_{12}}(\alpha_{12}
\beta_1\delta_{12}
- \alpha_{32} \gamma_1 k_1 + \gamma_1 k_1^2),\\
\phi^{(2)}&=&\frac{s_2E_2 }{\beta_1\delta_{12}}(\alpha_{12}
\beta_1\delta_{12}
- \alpha_{32} \gamma_1 k_2 +\gamma_1 k_2^2)+
\\\nonumber
&&\frac{s_3E_3 }{\beta_1\delta_{12}}(\alpha_{12} \beta_1
\delta_{12}
- \alpha_{32} \gamma_1 k_3 + \gamma_1 k_3^2),\\\nonumber
&&E_n=e^{k_n x +q_n y + \omega_n t+\nu_n \tau},\;\;n=1,2,3,
\end{eqnarray}
where
\begin{eqnarray}
\label{qomeganu}
q_n=\frac{\beta_2}{\beta_1} k_n,\;\;
\omega_n=-\frac{k^2_n}{\beta_1}(\beta_1+\beta_2 \gamma_1),
\;\;
\nu_n=-\frac{k_n^2
\beta_2}{\beta_1^2}(\beta_2+\beta_1 \gamma_2),\;\;\\\nonumber
n=1,2,3,
\end{eqnarray}
which are in agreement with eqs.(\ref{solpsi1})-(\ref{param1}).
Determinants (\ref{det1}) are expressed as follows:
\begin{eqnarray}\label{det01}
\triangle&=& \frac{s_1 E_1}{\beta_1} ((k_1-k_2) s_2 E_2 + 
(k_1-k_3) s_3 E_3),\\\label{det11}
\triangle_1&=& -\frac{\gamma_1 s_1 E_1}{\beta_1\delta_{12}} 
((k_1-k_2)(\alpha_{32} -k_1-k_2) s_2 E_2 +\\\nonumber 
 &&(k_1-k_3)(\alpha_{32} -k_1-k_3) s_3 E_3 ) ,\\\label{det21}
\triangle_2&=& \frac{ s_1 E_1}{\beta_1^2\delta_{12}} \times\\\nonumber
&&((k_1-k_2) (\alpha_{12} \beta_1\delta_{12} - 
\alpha_{32}^2 \gamma_1 + 
\alpha_{32} \gamma_1(k_1+k_2) - \gamma_1 k_1 k_2) s_2
E_2+\\\nonumber
&&(k_1-k_3) (\alpha_{12} \beta_1\delta_{12}  -
 \alpha_{32}^2 \gamma_1 + 
\alpha_{32} \gamma_1(k_1+k_3) - \gamma_1 k_1 k_3) s_3 E_3
).
\end{eqnarray}
Functions $u_1$ and $u_2$   have no singularities provided
that $\triangle\neq 0$ for all values of parameters $x,y,t,\tau$,
which happens if only
\begin{eqnarray}
{\mbox {sign}}((k_1-k_2) s_2)= {\mbox {sign}}((k_1-k_3) s_3).
\end{eqnarray}

Taking into account relations (\ref{param1}), system (\ref{eq1},\ref{eq2}) 
gets the following form:
\begin{eqnarray}\label{eq1n}
{u_1}_t + {u_1}_{xx} + \gamma_1 {u_1}_{xy} + 
\frac{\delta_{12}}{\beta_1\gamma_1}(2\beta_1+\beta_2\gamma_1) u_1
{u_1}_x+&&\\\nonumber
\delta_{12} u_1 {u_1}_y + 
(2 \beta_1+ \beta_2 \gamma_1) {u_2}_x +
 \beta_1 \gamma_{1} {u_2}_y&=&0,\;\;\\\label{eq2n}
{u_1}_\tau+{u_1}_{yy}+ \gamma_2 {u_1}_{xy} +
\frac{\gamma_2\delta_{12}\beta_2}{\gamma_1\beta_1}u_1 {u_1}_x+
\frac{\delta_{12}}{\gamma_1\beta_1} ( 2\beta_2 + \beta_1\gamma_2)
 u_1 {u_1}_y+&&\\\nonumber
\beta_2 \gamma_{2} {u_2}_x+
(2\beta_2+\beta_1\gamma_2 ) {u_2}_y&=&0,\;\;\\\nonumber
\end{eqnarray}

Note that owing to the relations (\ref{q}-\ref{nu}) functions $u_1$
and $u_2$ depend on two variables rather then four: 
\begin{eqnarray}\label{XT}
X=x+\frac{\beta_2}{\beta_1} y,\;\;T=\frac{\beta_1+\beta_2
\gamma_1}{\beta_1} t + \frac{\beta_2(\beta_2 + \beta_1
\gamma_2)}{\beta_1^2} \tau
\end{eqnarray}
Because of this fact, functions $u_1$ and $u_2$, given by the  
formulas (\ref{sol1},\ref{det01}-\ref{det21}),
represent single kink, which is
essentially one-dimensional.
Moreover, in virtue of eqs.(\ref{XT}), the above system
(\ref{eq1n},\ref{eq2n}) can be
written in the form of differentiated Burghers equation on the function $u_1$
 with independent
variables $X$ and $T$.

\subsubsection{Multidimensional kinks}

In this section $N=4$, superscript $i$ takes values from 1 to 4,
unless otherwise specified.
 The system (\ref{ALGSYSTN}) has the form 
\begin{eqnarray}\label{algsyst2}
\phi^{(i)}=u_1 \psi_1^{(i)} + u_2 \psi_2^{(i)} + 
u_3 \psi_3^{(i)} +  u_4 \psi_4^{(i)}.
\end{eqnarray}
Formulas (\ref{sol1}) for the functions
$u_1$ and $u_2$ are held with determinants in the following form:
\begin{eqnarray}\label{det2}
\triangle &=& \left|
\begin{array}{cccc}
\psi_1^{(1)}&\psi_2^{(1)}&\psi_3^{(1)}&\psi_4^{(1)}\cr
\psi_1^{(2)}&\psi_2^{(2)}&\psi_3^{(2)}&\psi_4^{(2)}\cr
\psi_1^{(3)}&\psi_2^{(3)}&\psi_3^{(3)}&\psi_4^{(3)}\cr
\psi_1^{(4)}&\psi_2^{(4)}&\psi_3^{(4)}&\psi_4^{(4)}
\end{array}
\right|,\;\;
\triangle_1 \;=\; \left|
\begin{array}{cccc}
\phi^{(1)}&\psi_2^{(1)}&\psi_3^{(1)}&\psi_4^{(1)}\cr
\phi^{(2)}&\psi_2^{(2)}&\psi_3^{(2)}&\psi_4^{(2)}\cr
\phi^{(3)}&\psi_2^{(3)}&\psi_3^{(3)}&\psi_4^{(3)}\cr
\phi^{(4)}&\psi_2^{(4)}&\psi_3^{(4)}&\psi_4^{(4)}
\end{array}
\right|
,\;\;\\\nonumber
\triangle_2 &=& \left|
\begin{array}{cccc}
\psi_1^{(1)}&\phi^{(1)}&\psi_3^{(1)}&\psi_4^{(1)}\cr
\psi_1^{(2)}&\phi^{(2)}&\psi_3^{(2)}&\psi_4^{(2)}\cr
\psi_1^{(3)}&\phi^{(3)}&\psi_3^{(3)}&\psi_4^{(3)}\cr
\psi_1^{(4)}&\phi^{(4)}&\psi_3^{(4)}&\psi_4^{(4)}
\end{array}\right|
\end{eqnarray}
Functions $\psi_1^{(i)}$, $\psi_2^{(i)}$, $\phi^{(i)}$ are
defined by the same formulas
(\ref{phi1n}-\ref{psi2n}) and (\ref{psiphiLn})
($j=1,2$) together with solutions (\ref{solpsi1})-(\ref{param1}). 
Because of
this, solution (\ref{sol1}) may  depend
 on four arbitrary functions of
single variable (compare with the paragraph below the eq.(\ref{param1})).

Functions $\psi_3^{(i)}$ and $\psi_4^{(i)}$  are solutions of the
equations (\ref{psijn},\ref{psiphiLn}) with $j=3,4$:
\begin{eqnarray}\label{geneq1}
{\psi_3^{(i)}}_x=\alpha_{53} \psi_3^{(i)}+ 
\alpha_{54} \psi_4^{(i)}+ \alpha_{52}
\psi_2^{(i)},\;\;{\psi_3^{(i)}}_y=\alpha_{63} \psi_3^{(i)}+ 
\alpha_{64} \psi_4^{(i)}+ \alpha_{62}
\psi_2^{(i)},\\\label{geneq2}
{\psi_4^{(i)}}_x=\alpha_{73} \psi_3^{(i)}+ 
\alpha_{74} \psi_4^{(i)}+ \alpha_{72}
\psi_2^{(i)},\;\;
{\psi_4^{(i)}}_y=\alpha_{83} \psi_3^{(i)}+ 
\alpha_{84} \psi_4^{(i)}+ \alpha_{82}
\psi_2^{(i)},\\
\label{geneq3}
l_k \psi^{(i)}_j=0,\;\;j=3,4,\;\;k=1,2.{\mbox{\hspace{5cm}}}
\end{eqnarray}
This  is a system of linear nonhomogeneous 
equations with solution in  the following general form:
\begin{eqnarray}\label{sol2}
\psi_3^{(i)}=\psi_{30}^{(i)} + \psi_{3p}^{(i)},\;\;
\psi_4^{(i)}=\psi_{40}^{(i)} + \psi_{4p}^{(i)},\;\;
\end{eqnarray}
where $\psi_{j0}^{(i)}$  and $\psi_{jp}^{(i)}$ 
are general solution of homogeneous system, related with the
system (\ref{geneq1}-\ref{geneq3}), and particular solution  of
nonhomogeneous system (\ref{geneq1}-\ref{geneq3}) respectively.

The general solution of homogeneous  system reads:
\begin{eqnarray}\label{psi34n}
\psi_{30}^{(i)}&=& p_1^{(i)} r_{3}^{(1)} 
e^{\tilde k_1 x + \tilde q_1 y +\tilde \omega_1 t +\tilde \nu_1 \tau}+
p_2^{(i)} r_{3}^{(2)} 
e^{\tilde k_2 x + \tilde q_2 y +\tilde \omega_2 t +\tilde \nu_2 \tau},\\
\psi_{40}^{(i)}&=& p_1^{(i)} r_{4}^{(1)} 
e^{\tilde k_1 x + \tilde q_1 y +\tilde \omega_1 t +\tilde \nu_1 \tau}+
p_2^{(i)} r_{4}^{(2)} 
e^{\tilde k_2 x + \tilde q_2 y +\tilde \omega_2 t +\tilde \nu_2
\tau},\\\label{ri}
r_{4}^{(i)}&=&\frac{(\tilde k_i-\alpha_{53})
r_{3}^{(i)}}{\alpha_{54}},\;\;\\\label{tildeq}
\tilde q_j&=&\frac{1}{\alpha_{54}}(\alpha_{74} \tilde k_j +
\alpha_{54}\alpha_{73} -
\alpha_{53}\alpha_{74}),\;\;j=1,2,\\\label{tildek}
\tilde k_{1,2}&=&\frac{1}{2}\left(
\alpha_{53} + \alpha_{64} \pm
\right.\\\nonumber
&&\left.
\frac{1}{\sqrt{\alpha_{74}}}
\sqrt{\alpha_{53}^2 \alpha_{74}-2 \alpha_{53} \alpha_{64}
\alpha_{74} +
\alpha_{64}^2 \alpha_{74} + 4 \alpha_{54}^2 \alpha_{83}}
\right),
\end{eqnarray}
where $\tilde \omega_j $ and $\tilde \nu_j$ are related with  $\tilde k_j$
and $\tilde q_j$ by the equations
\begin{eqnarray}\label{tildeomeganu}
\tilde \omega_j=-\tilde k^2_j- \tilde k_j \tilde q_j \gamma_1,\;\;
\tilde \nu_j = -\tilde q^2_j - \tilde k_j \tilde q_j \gamma_2,
\end{eqnarray}
 which follows from the eqs.(\ref{geneq3}).

The
particular solution
can be taken in the form (in virtue of the eq.(\ref{solpsi2})) 
\begin{eqnarray}\label{partsol}
\psi_{3p}^{(i)}&=&\int \Gamma_1^{(i)}(k) c_2^{(i)}(k) \exp((x+\beta_2/\beta_1 y)
k + \omega(k) t + \nu(k) \tau) dk,\\\nonumber
\psi_{4p}^{(i)}&=&\int \Gamma_2^{(i)}(k) c_2^{(i)}(k) \exp((x+\beta_2/\beta_1 y)
k + \omega(k) t + \nu(k) \tau) dk,
\end{eqnarray}
with $\omega(k)$  and $\nu(k)$ given by the
eq.(\ref{omega},\ref{nu}).
Substitution  of the eqs.(\ref{partsol}) into the
eqs.(\ref{geneq1},\ref{geneq2}) results in the following 
expressions for $\Gamma_1$ and $\Gamma_2$ together 
with additional relations among 
parameters $\alpha_{ij}$:
\begin{eqnarray}\label{Gamma}
\Gamma_1(k)&=&\frac{\alpha_{52}^2}{\alpha_{52} k - \alpha_{52}
\alpha_{53} - \alpha_{54}\alpha_{62}},\;\;
\Gamma_2(k)\;=\;\frac{\alpha_{62}}{\alpha_{52}}
\Gamma_1(k),\\\nonumber
\alpha_{52}&=&\frac{\alpha_{54}\alpha_{62}
(\alpha_{54}\beta_2-\alpha_{74} \beta_1)}{
(\alpha_{53} \alpha_{74}\beta_1 - \alpha_{64}
\alpha_{74}\beta_1+
\alpha_{54} \alpha_{84}\beta_1
-\alpha_{53}\alpha_{54}\beta_2)},\\
\alpha_{72}&=&\frac{\alpha_{52}\beta_2}{\beta_1} ,\;\;
\alpha_{82}\;=\;\frac{\alpha_{62}\beta_2}{\beta_1},\\
\alpha_{83}&=&\frac{
\alpha_{74} (\alpha_{84}\beta_1-\alpha_{74}
\beta_2)}{\alpha_{54}(\alpha_{54}\beta_2-\alpha_{74} \beta_1)^2}
\times\\\nonumber
&&
(\alpha_{53} \alpha_{74}\beta_1 - \alpha_{64}
\alpha_{74}\beta_1+
\alpha_{54} \alpha_{84}\beta_1
-\alpha_{53}\alpha_{54}\beta_2)
\end{eqnarray}

As an example, let us choose the following expressions
for $\psi_j^{(i)}$ and $\phi^{(i)}$:
\begin{eqnarray}
\psi_1^{(i)} &=&s_i c_1^{(i)} e^{k_i x+q_i y + \omega_i t + \nu_i
\tau},\;\;i=1,2,3,\\
\psi_1^{(4)} &=&s_4 c_1^{(4)} e^{k_4 x+q_4 y + \omega_4 t + \nu_4
\tau}+s_5 c_1^{(5)} e^{k_5 x+q_5 y + \omega_5 t + \nu_5 \tau},\\
\psi_2^{(i)} &=&s_i  e^{k_i x+q_i y + \omega_i t + \nu_i
\tau},\;\;i=1,2,3,\\
\psi_2^{(4)} &=&s_4  e^{k_4 x+q_4 y + \omega_4 t + \nu_4
\tau}+s_5  e^{k_5 x+q_5 y + \omega_5 t + \nu_5 \tau},\\
\phi^{(i)} &=&s_i c^{(i)} e^{k_i x+q_i y + \omega_i t + \nu_i
\tau},\;\;i=1,2,3,\\
\phi^{(4)} &=&s_4 c^{(4)} e^{k_4 x+q_4 y + \omega_4 t + \nu_4
\tau}+s_5 c^{(5)} e^{k_5 x+q_5 y + \omega_5 t + \nu_5 \tau},\\
\psi_3^{(1)}&=& p_1  e^{\tilde k_1 x + \tilde q_1 y 
+ \tilde \omega_1 t + \tilde \nu_1 \tau} +
\Gamma_1(k_1) s_1  e^{k_1 x+q_1 y + \omega_1 t + \nu_1
\tau},\\
\psi_3^{(2)}&=& p_2  e^{\tilde k_2 x + \tilde q_2 y 
+ \tilde \omega_2 t + \tilde \nu_2 \tau} +
\Gamma_1(k_2) s_2  e^{k_2 x+q_2 y + \omega_2 t + \nu_2
\tau},\\
\psi_3^{(3)}&=& 
\Gamma_1(k_3) s_3  e^{k_3 x+q_3 y + \omega_3 t + \nu_3
\tau},\\
\psi_3^{(4)}&=& 
\Gamma_1(k_4) s_4  e^{k_4 x+q_4 y + \omega_4 t + \nu_4
\tau}+\Gamma_1(k_5) s_5  e^{k_5 x+q_5 y + \omega_5 t + \nu_5
\tau},\\
\psi_4^{(1)}&=& p_1 r_4^{(1)} e^{\tilde k_1 x + \tilde q_1 y 
+ \tilde \omega_1 t + \tilde \nu_1 \tau} +
\Gamma_2(k_1) s_1  e^{k_1 x+q_1 y + \omega_1 t + \nu_1
\tau},\\
\psi_4^{(2)}&=& p_2 r_4^{(2)} e^{\tilde k_2 x + \tilde q_2 y 
+ \tilde \omega_2 t + \tilde \nu_2 \tau} +
\Gamma_2(k_2) s_2 e^{k_2 x+q_2 y + \omega_2 t + \nu_2
\tau},\\
\psi_4^{(3)}&=& 
\Gamma_2(k_3) s_3  e^{k_3 x+q_3 y + \omega_3 t + \nu_3
\tau},\\
\psi_4^{(4)}&=& 
\Gamma_2(k_4) s_4  e^{k_4 x+q_4 y + \omega_4 t + \nu_4
\tau}+\Gamma_2(k_5) s_5  e^{k_5 x+q_5 y + \omega_5 t + \nu_5
\tau}.
\end{eqnarray}
where $c^{(n)}\equiv c^{(n)} (k_n) $, 
$c_1^{(n)}\equiv c_1^{(n)} (k_n) $ 
are given by the eqs.(\ref{ci},\ref{c1i}) with $c_2^{(n)}=1$;
$r_4^{(n)}$ 
are given by the eq.(\ref{ri}) with $r_3^{(n)}=1$;
parameters $k_n,q_n,\omega_n,\nu_n$ are mutually related by the
eqs.(\ref{q}-\ref{nu}); relationship among 
parameters $\tilde k_n,\tilde q_n,\tilde \omega_n,\tilde \nu_n$ 
is given  by the eqs.(\ref{tildeq}-\ref{tildeomeganu});
$\Gamma_n$ are given by the eqs.(\ref{Gamma});
$s_n$ and  $p_n$ are arbitrary parameters.

General expressions for the 
functions $u_1$ and $u_2$ are very
complicated. 
As an example, 
let us fix the following list of parameters:
\begin{eqnarray}
k_1&=&1,\;\;k_2\;=\;-1,\;\;k_3\;=\;2,\;\;k_4\;=\;-2,\;\;k_5\;=\;0,\\
\beta_1&=&1,\;\;\beta_2\;=\;-3,\;\;\gamma_1\;=\;\gamma_2\;=\;\delta_{12}\;=\;1,\\
\alpha_{12}&=&\alpha_{32}\;=\;\alpha_{53}\;=\;\alpha_{62}
\;=\;\alpha_{74}\;=\;1,\\
\alpha_{54}&=&\alpha_{84}\;=\;2,\;\;\alpha_{64}\;=\;-1.
\end{eqnarray}
Then, formulas (\ref{sol1}) read:
\begin{eqnarray}\label{u1}
u_1&=&\frac{45}{\tilde \triangle}
\left(s_5 e^{2( x +  12 \tau)}-2 s_4 e^{2(3 y +4
t)}\right)\times\\\nonumber
&&
\left( 
343 s_2 e^{\frac{6}{7} y+\frac{48}{49} t} + 76 p_2
e^{\frac{2}{7}x+\frac{144}{49} \tau}
\right),\\\label{u2}
u_2&=&\frac{6}{\tilde\triangle}
\left(15680 s_2s_4 e^{\frac{48}{7}  y+\frac{440}{49}t} +8967 s_2
s_5 
e^{ 2 x +\frac{6}{7} y + \frac{48}{49} t + 24 \tau}
 +\right.\\\nonumber
&&\left. 4560  s_4p_2 e^{\frac{2}{7} x + 6 y + 8 t +
\frac{144}{49} \tau} +1140 s_5 p_2 
e^{\frac{16}{7} x + \frac{1320}{49} \tau}
\right)
,\\\nonumber
&&\tilde\triangle=24010 s_2 s_4 e^{ \frac{48}{7} 
y+\frac{440}{49}t} +
21609 s_2 s_5 e^{2 x +\frac{6}{7} y + \frac{48}{49} t + 24 \tau }
+\\\nonumber
&& 
6840 s_4 p_2 e^{ \frac{2}{7} x + 6 y + 8 t +
\frac{144}{49} \tau}  + 3420 s_5p_2 e^{ \frac{16}{7} x + \frac{1320}{49} \tau}.
\end{eqnarray}
This solution has no singularities if
\begin{eqnarray}
{\mbox{sign}}(s_2s_4)={\mbox{sign}}(s_2s_5)=
{\mbox{sign}}(s_4p_2)={\mbox{sign}}(s_5p_2)
\end{eqnarray}
The nonlinear system keeps the same form (\ref{eq1n},\ref{eq2n}).
The change of variables (\ref{XT}) is not effective in this
case, so that the nonlinear system is essentially (2+2)-dimensional.
If one of parameters $x$, $y$, $t$ or $\tau$ is fixed, then
functions $u_1$ and $u_2$ represent (2+1)-dimensional inelastic 
3-kinks interaction. 

\section{Conclusions}

The represented method  is another way  of using the algebraic
system of equations for  analysis of nonlinear PDE. In comparison with
method considered in \cite{Z1,Z2,Z3}, it supplies wide class of
particular solutions for nonlinear PDE which may be solved by
this method. In general, nonlinear PDE considered in this paper,
are not completely integrable in classical sence. For instance,
attempt to
introduce commuting flows by the standard way leads, generally
speaking,
 to new
constraints on the manifold of available solutions to nonlinear PDE. 

There is no formal restriction on the linear differential operators
$L_1$ and $L_2$ (see eqs.(\ref{LL})). This makes 
the algorithm  fluxible, but its application field is not
defined yet. An interesting problem is construction of commuting
flows on  submanifold of particular solutions to the given
nonlinear PDE.

\end{document}